\documentclass[letterpaper, 10 pt, conference]{ieeeconf}
\usepackage[utf8]{inputenc}

\IEEEoverridecommandlockouts    %used for \thanks
\usepackage{blindtext}

\usepackage{amsmath} % general nice math
\usepackage{bbm}
\usepackage{amssymb}
\usepackage{mathtools}
\usepackage{amsthm}
\usepackage[ruled]{algorithm2e}

\usepackage{booktabs}

\theoremstyle{definition}
\newtheorem{definition}{Definition}

\newtheorem{assumption}{Assumption}

\newtheorem{problem}{Problem}

\usepackage{cite}
\usepackage{comment}

\usepackage{todonotes}

\usepackage{todonotes}

\makeatletter
\let\NAT@parse\undefined
\makeatother
\usepackage{hyperref}
\hypersetup{
   colorlinks=true,
    linkcolor= blue,
    allcolors=blue,
    citecolor = blue,
    filecolor=black,      
    urlcolor=blue,
    }
\title{\Large \bf A Scalable Last-Mile Delivery Service: From Simulation to Scaled Experiment}
\author{Meera Ratnagiri$^\dagger$, Clare O'Dwyer$^\dagger$, Logan E. Beaver, \emph{IEEE Student Member},\\
Heeseung Bang, \emph{IEEE Student Member}, Behdad Chalaki, \emph{IEEE Student Member}, \\ Andreas A. Malikopoulos, \emph{IEEE Senior Member}, %
\thanks{This research was supported by the Sociotechnical Systems Center (SSC) at the University of Delaware.}%
\thanks{$\dagger$ These authors contributed equally to this work}%
\thanks{M. Ratnagiri is with Concord High School, Wilmington, DE 18910 USA}%
\thanks{C. O'Dwyer is with Archmere Academy, Claymont, DE 19703 USA}%
\thanks{L.E. Beaver, H. Bang, B. Chalaki, and A.A. Malikopoulos are with the Department of Mechanical Engineering, University of Delaware, Newark, DE 19716 USA (emails: \texttt{\{lebeaver;heeseung;bchalaki;andreas\}@udel.edu})}
}

\begin{document}

\maketitle

\begin{abstract}
In this paper, we investigate the problem of a last-mile delivery service that selects up to $N$ available vehicles to deliver $M$ packages from a centralized depot to $M$ delivery locations.
The objective of the last-mile delivery service is to jointly maximize customer satisfaction (minimize delivery time) and minimize operating cost (minimize total travel time) by selecting the optimal number of vehicles to perform the deliveries.
We model this as an assignment (vehicles to packages) and path planning (determining the delivery order and route) problem, which is equivalent to the NP-hard multiple traveling salesperson problem.
We propose a scalable heuristic algorithm, which sacrifices some optimality to achieve a reasonable computational cost for a high number of packages.
The algorithm combines hierarchical clustering with a greedy search.
To validate our approach, we compare the results of our simulation to experiments in a $1$:$25$ scale robotic testbed for future mobility systems. 
\end{abstract}

\section{Introduction}
% !!!! add a paragraph to the introduction and update the FIGURE 2-  DO NOT FORGET, !!!!!
In a rapidly urbanizing world, we need to make fundamental transformations in how we use and access transportation \cite{zhao2019enhanced}.
With the meteoric rise of the e-commerce industry, last-mile delivery, especially parcel delivery, has attracted considerable attention \cite{Hu2016}. 
Last-mile delivery refers to the last step in a supply chain, where goods or people are transported from a centralized hub to their final destination.
In general, last-mile delivery is considered the least efficient part of the entire logistics chain, and it has been an area of significant research \cite{lin2014survey,Huizing2015,Wang2016,yao2015vehicle,jiang2019travelling,wang2019car4pac,osaba2018multi}.
Additionally, as our energy, transportation, and cyber networks integrate further, and interact with human operators, we are witnessing a new level of complexity \cite{Malikopoulos2016c} in transportation systems.
In this environment, last-mile delivery firms will need to meet ever-increasing fulfillment demands as efficiently as possible.

This paper investigates the problem of last-mile delivery through a centralized delivery service, which we model as a joint assignment (assigning vehicles to packages) and path planning (determining delivery order and route) problem with a variable number of vehicles.
These problems are coupled, as the cost of the assignment is dependent on the path taken by each of the vehicles.
This is similar to the multiple traveling salesperson problem (mTSP) \cite{Baranwal2017,Bektas2006}, a generalization of the NP-hard traveling salesperson problem (TSP).
As a result, there have been many approaches to generating solutions to mTSP for vehicle routing and last-mile delivery problems \cite{Wang2020,al2019comparative,lu2016applying,liu2018research,Huizing2015,bramel1995location}.

Recent efforts in this area include ant colony optimization to solve mTSP with capacity and time window constraints for vehicle routing \cite{Wang2020}, fuzzy logic approaches for multi-objective mTSP  \cite{trigui2017fl}, and data-driven graph-theoretic approaches to bus scheduling \cite{Tong2019}.
Other efforts have proposed augmenting last-mile delivery with cargo bicycles \cite{naumov2021identifying} and drones \cite{remer2019multi,di2020trucks} to decrease the number of cars and vans required and further reduce delivery times.
However, to the best of our knowledge, validation of a last-mile delivery service in a physical experiment has not yet been reported in the literature.
We seek to fill this gap by validating a heuristic algorithm for last-mile delivery in a physical scaled testbed.

Our first contribution is a fast online algorithm to generate the assignment of vehicles to deliveries as well as the routes of each vehicle.
This is based on our previous work, where agents are assigned to goals and generate trajectories to create a desired formation \cite{Beaver2019AGeneration,bang2021energy}.
%In addition, last-mile delivery has been extensively studied using data-driven simulations for static and dynamic settings \cite{bertsimas2019optimizing,Tong2019}.
%However, to the best of our knowled examples of algorithms being implemented in physical transportation systems.
Our second contribution is the experimental validation of a last-mile delivery algorithm  in a physical testing environment, which, to the best of our knowledge, has not yet been addressed in the literature.

The remainder of the paper is organized as follows. In Section \ref{sec:prelims}, we present some mathematical preliminaries on graph theory, and in Section \ref{sec:problem}, we formulate the last-mile delivery service problem along with our modeling framework.
In Section \ref{sec:solution}, we discuss our scalable heuristic solution, and in Section \ref{sec:simulation}, we present simulation results alongside the experimental validation in our $1$:$25$ scale testbed.
%We present experimental validation in our $1$:$25$ scale testbed in Section \ref{sec:experiment}, and f
Finally, we draw concluding remarks in Section \ref{sec:conclusion}.

\section{Preliminaries}\label{sec:prelims}
In this section, we present preliminary mathematical material that describes how we model our urban road network as a mathematical graph.
\begin{definition}\label{def:graph}
Our urban roadway network is represented by a directed graph, denoted $\mathcal{U} = \big(\mathcal{V}, \mathcal{E}\big)$, which consists of:
\begin{itemize}
    \item A set of nodes, $\mathcal{V}\subset\mathbb{N}$, corresponding to the locations where 1) two road segments join or separate, e.g., merging zones, intersections, roundabouts, entry and exit ramps, and 2) locations where the road network transitions between straight-line and arc segments.
    \item A set of directed edges, $\mathcal{E}\subset\mathcal{V}\times\mathcal{V}$, which contains pairs of nodes that are connected by a one-way road segment.
\end{itemize}
%\todoHeeseung{Do we actually use this $e$ as an edge later?}
\end{definition}
For every edge $e\in\mathcal{E}$ there is an associated cost, $c(e)\in\mathbb{R}_{> 0},$ which corresponds to the amount of time taken for a vehicle to travel the length of the edge.
Next, we define the motion of vehicles through the road network in terms of paths.

\begin{definition} \label{def:path}
For a graph $\mathcal{U} = \big(\mathcal{V}, \mathcal{E}\big)$, a path of length $\ell\in\mathbb{N}$ is a sequence of nodes, denoted  $\mathcal{P}=\left(p_1,\dots,p_{\ell}\right)$, where $p_1,\dots,p_{\ell}\in\mathcal{V}$, and there exists a corresponding edge $(p_k, p_{k+1})\in\mathcal{E}$ for all $k\in\{1,\dots,\ell-1\}$.
The total cost of a path $\mathcal{P}$ is equal to the sum of the costs of its edges, i.e., $c\left(\mathcal{P}\right) = \sum_{k=1}^{\ell-1} c\big( \left(p_k, p_{k+1}\right) \big)$.
%$p_k\in\mathcal{V}$ for all $k\in \{1,\dots,\ell\}$ and there exists a corresponding edge $(p_k, p_{k+1})\in \mathcal{E}$ in the graph $\mathcal{G}$ for all $k\in\{1,\dots,\ell-1\}$.

%$\mathcal{P}\subset\mathbb{V}$ is a sequence of nodes such that for any two sequential elements $(p_k)$, $(p_{k+1}) \in\mathcal{P}$ there exists a corresponding edge $(p_k, p_{k+1}) \in \mathcal{E}$ in the graph $\mathcal{G}$.
\end{definition}

In the next section, we formulate our last-mile delivery service, which manages the delivery of packages to customers from a centralized depot.
The objective of our delivery service is to jointly minimize cost and maximize customer satisfaction, i.e., jointly minimize the total travel time and average delivery time.

%the problem of a delivery service that must deliver packages to passengers from a centralized depot while simultaneously minimizing travel cost and maximizing customer satisfaction.

\section{Problem Formulation}\label{sec:problem}
Our delivery service consists of $N\in\mathbb{N}$ vehicles that can be used to deliver $M\in\mathbb{N}$ packages to nodes located on the urban road network $\mathcal{U}$.
Let $\mathcal{N}\coloneqq\{1,\dots,N\}$ index the available delivery vehicles, and let $\mathcal{D} \subset \mathcal{V}$ denote the set of all delivery locations. %points.
Finally, let $\mathcal{G} = \{1, 2, \dots, M\}$ uniquely index the delivery locations in $\mathcal{D}$, i.e., there exists a bijective mapping
\begin{equation} \label{eq:map}
    m : \mathcal{D} \to \mathcal{G}.
\end{equation}

To formulate our delivery service problem, the first step is to partition the delivery locations into, at most, $N$ disjoint sets that the vehicles can be assigned to.
We achieve this by defining an assignment matrix.
\begin{definition}\label{def:assignment}
The assignment matrix, $\mathbf{A}$, is an $N \times M$ binary matrix, where element $a_{ij} \in \{0, 1\}$ is $1$ if vehicle $i\in\mathcal{N}$ is assigned to deliver a package to delivery location $j\in\mathcal{G}$ and $0$ otherwise.
\end{definition}

%\todoHeeseung{Although nemerically they are same, writing 'deliver package $j\in\mathcal{G}$' seems a bit weird as we defined $\mathcal{G}$ as am index set of the delivery locations. How do you guys think?}

The objective of our delivery service is to determine the number of vehicles to send on deliveries such that operating cost and customer satisfaction are jointly optimized.
We model customer satisfaction as being inversely proportional to package delivery time, with the cost denoted by $J_s$, and we denote the total travel cost by $J_c$, which is proportional to the total round-trip time of all vehicles, i.e.,
\begin{align}
    J_s(\mathbf{A}) &= \sum_{j\in\mathcal{G}} {\frac{t_j(\mathbf{A})}{M}}, \label{eq:satisfaction} \\
    J_c(\mathbf{A}) &= \sum_{i\in\mathcal{N}}{T_i(\mathbf{A})}, \label{eq:final-time}
\end{align}
where $t_j$ denotes the time taken for package $j\in\mathcal{G}$ to be delivered, and $T_i\in\mathbb{R}_{\geq 0}$ is the round-trip travel time for vehicle $i$ to make its deliveries and return to the depot. 
Note that both $T_i$ and $t_j$ depend on the actual route taken by the vehicles, which is a function of the assignment matrix, $\mathbf{A}$.
We determine the assignment matrix, and thus the optimal number of vehicles to deploy, by solving the following centralized assignment problem.
\begin{problem}[Delivery Assignment] \label{prb:assignment}
Assign vehicles to deliver packages such that the average delivery time and round-trip travel time are jointly minimized,
	\begin{align}
	\min_{\mathbf{A}} & \Bigg\{ \alpha J_s(\mathbf{A}) + (1-\alpha)J_c(\mathbf{A}) \Bigg\} \label{eq:totaltravelCost}
 	\\
 	\text{subject to:}&\notag\\
 	\sum_{i\in\mathcal{N}} a_{ij}= 1 &\text{ for all } j\in\mathcal{G} \notag, \\
 	a_{ij} \in \{0, 1\} &\text{ for all } i\in\mathcal{N}, j\in\mathcal{G}\notag,
     \end{align}
    where $0 \leq \alpha \leq 1$ is an intrinsic parameter that balances the tradeoff between minimizing operating cost and maximizing customer satisfaction for a particular delivery service, and $a_{ij},~ i\in\mathcal{N},~ j\in\mathcal{G},$ are the elements of the assignment matrix $\mathbf{A}.$
    Note that the first constraint ensures that each package gets delivered only by one vehicle, and some vehicles may not be assigned to deliver any packages. 
\end{problem}

In order to deliver the packages, and therefore compute the cost components of Problem \ref{prb:assignment}, we need to determine the route taken by each vehicle given the assignment matrix, $\mathbf{A}$.
We define the delivery route of vehicle $i\in\mathcal{N}$ as the sequence $\mathcal{S}_i \coloneqq \left(s_1, s_2,\dots, s_n\right)$, where $n = \sum_{j\in\mathcal{G}}~{a_{ij}}$ (Definition \ref{def:assignment}) and $\bigcup_{k\in\{1,\dots,n\}} s_k = \{ v \in \mathcal{D} ~|~ j = m(v), \, a_{ij} = 1 \}$,
where $m$ maps delivery locations to indices by \eqref{eq:map}.
To determine the optimal sequence to deliver their assigned packages, each vehicle $i$ solves the following problem.

\begin{problem}\label{prb:path}
For each vehicle $i\in\mathcal{N}$, with a given delivery assignment matrix $\mathbf{A}$, determine the path $\mathcal{P}_i$ that minimizes the total travel time such that the vehicle starts and ends at the depot and delivers all assigned packages, i.e., 
\begin{align}
    \min_{\mathcal{P}_i, \mathcal{S}_i} ~&~ T_i\\
    \text{subject to:}& \notag\\
    p_1 &= d, \quad p_1 \in \mathcal{P}_i ,\notag\\
    p_{\ell} &= d, \quad \ell = |\mathcal{P}_i|,\,\,p_{\ell}\in\mathcal{P}_i, \notag\\
    \mathcal{S}_i &\text{ is a subsequence of } \mathcal{P}_i\nonumber,
\end{align}
%%%
where $\mathcal{P}_i = \left(p_1, p_2, \dots, p_{\ell}\right)$  is the path taken by vehicle $i$ (Definition \ref{def:path}), $\mathcal{S}_i = \left(s_1, s_2,\dots, s_n\right)$ defines the sequence of deliveries, and $d\in\mathcal{V}$ is the node corresponding to the depot entrance.
\end{problem}

Thus, the cost of the optimal assignment in Problem \ref{prb:assignment} is determined by each vehicle's solution to Problem \ref{prb:path}, which is equivalent to the TSP.
In the following section we detail our scalable solution to solve Problems \ref{prb:assignment} and \ref{prb:path} in simulation, and we verifying this solution experimentally.
To this end, we impose the following assumptions.
\begin{assumption} \label{smp:automation}
The delivery vehicles are equipped with connected and automated vehicle technologies that significantly reduces the effects of traffic bottlenecks and delays \cite{Mahbub2019ACC,chalaki2020TCST}.
\end{assumption}
\begin{assumption}\label{smp:speedLimits}
The speed limits on the road network are fixed and known a priori.
\end{assumption}

Assumptions \ref{smp:automation} and \ref{smp:speedLimits} ensure that the environment is deterministic when the vehicles plan their routes.
These assumptions can be relaxed by including a time-varying term to the edge cost to account for traffic lights, stop signs, and congestion.
Assumption \ref{smp:automation} can also be relaxed by allowing different delivery modes that bypass congestion, e.g., motorcycle, bicycle, and drone delivery.
In the case that Assumption \ref{smp:speedLimits} is relaxed, and the traffic information is stochastic, a data-driven approach could be used to calculate the expected delay at each edge.

\begin{assumption}\label{smp:distanceLimits}
The vehicles have sufficient energy to achieve their assigned delivery routes.
\end{assumption}
\begin{assumption}\label{smp:delivery}
The delivery locations are fixed and known a priori and without constraints on the delivery time.
\end{assumption}

Assumptions \ref{smp:distanceLimits} and \ref{smp:delivery} ensure that any vehicle can be assigned to any sequence of deliveries, and that a solution is always guaranteed to exists to Problem \ref{prb:assignment}.
Assumption \ref{smp:distanceLimits} can be relaxed by including a total energy cost constraint in Problem \ref{prb:path} and discarding any assignments that impose infeasible delivery routes. 
Assumption \ref{smp:delivery} is relevant for last-mile delivery, where the delivery service knows what packages must be shipped out before they arrive.
This assumption can be relaxed by adding a delivery time constraint to Problem \ref{prb:path} and discarding any assignments that don't satisfy the delivery constraints. 
If either assumption is relaxed, then it is necessary to derive conditions for the existence of a feasible assignment matrix.

\section{Solution Approach}\label{sec:solution}

In this section, we describe a scalable method of assigning vehicles to packages, and determining their delivery routes, in order to minimize the joint travel and delivery time costs.
In Problem \ref{prb:assignment}, the parameter $\alpha$ describes the structure of a particular company and how the cost of vehicle usage and labor is weighed against the benefits of faster delivery times. A higher $\alpha$ implies that $J_s(\mathbf{A})$ is the dominant term, and this will result in solutions that use more vehicles to decrease average delivery time.
In contrast, a small value of $\alpha$ implies that  $J_c(\mathbf{A})$ dominates, and this will result in fewer delivery vehicles being assigned to reduce the total amount of time spent traveling across all vehicles.

First, our delivery service receives the number of packages, $M$, and their delivery locations.
To solve the assignment of vehicles to packages (Problem \ref{prb:assignment}), we employ complete-linkage clustering \cite{dawyndt2005complete}.
While complete-linkage clustering may not always yield the optimal solution to Problem \ref{prb:assignment}, it is a hierarchical clustering algorithm.
Therefore, it is able to generate an approximately optimal assignment of vehicles to packages for every case simultaneously, i.e., assignments for using $1, 2, \dots, N$ vehicles, in $O(n^2)$.
Complete-linkage clustering also ensures that the maximum distance between any two packages in the same cluster is minimized.
In contrast, to find the optimal assignment, the Hungarian Algorithm has a computational complexity of $O(n^3)$ for a single scenario, i.e., a fixed number of participating vehicles.
This requires significantly more evaluations of Problem \ref{prb:path}, thus we selected the complete-linkage clustering algorithm to ensure scalability.

In complete-linkage clustering, the $M$ delivery locations are first clustered into $M$ groups, i.e., each delivery assignment is a singleton.
This corresponds to the $M$ leaf nodes in our hierarchical clustering tree.
Next, the two packages with the shortest separating distance are combined, resulting in $M-1$ groups of packages, which adds a branch to our hierarchical clustering tree.
This process is repeated until only a single group, containing all $M$ packages, remains; this corresponds to the root node of our hierarchical clustering tree.
Finally, we start at the root node and travel along the hierarchical clustering tree to determine the assignment of $1, 2, \dots, N$ vehicles to the $M$ packages.
Thus, we can enumerate all of the resulting $1, 2, \dots, N$ assignments to determine which one minimizes the total cost.

Problem \ref{prb:path} is a modified TSP, which contains additional nodes that are not associated with a package delivery.
These problems are known to be NP-hard, with a computational cost approaching $O(n!)$.
Thus, to ensure our approach scales with a large number of vehicles, we solve Problem \ref{prb:path} using a greedy search algorithm.
Each vehicle $i\in\mathcal{N}$ begins at the depot, and it calculates the shortest path from its current location to each of its assigned delivery locations.
The vehicle selects the nearest delivery location as the first element of its delivery sequence $\mathcal{S}_i$, and the second element is selected by comparing the distance of all remaining packages to $s_1 \in \mathcal{S}_i$.
This process is repeated until $\mathcal{S}_i$ contains all delivery locations exactly once, and this also yields the path $\mathcal{P}_i$ taken by the vehicle.
In the case of ties, the package with the lower index is arbitrarily selected.

\begin{figure}[ht]
    \centering
    \includegraphics[width=\linewidth]{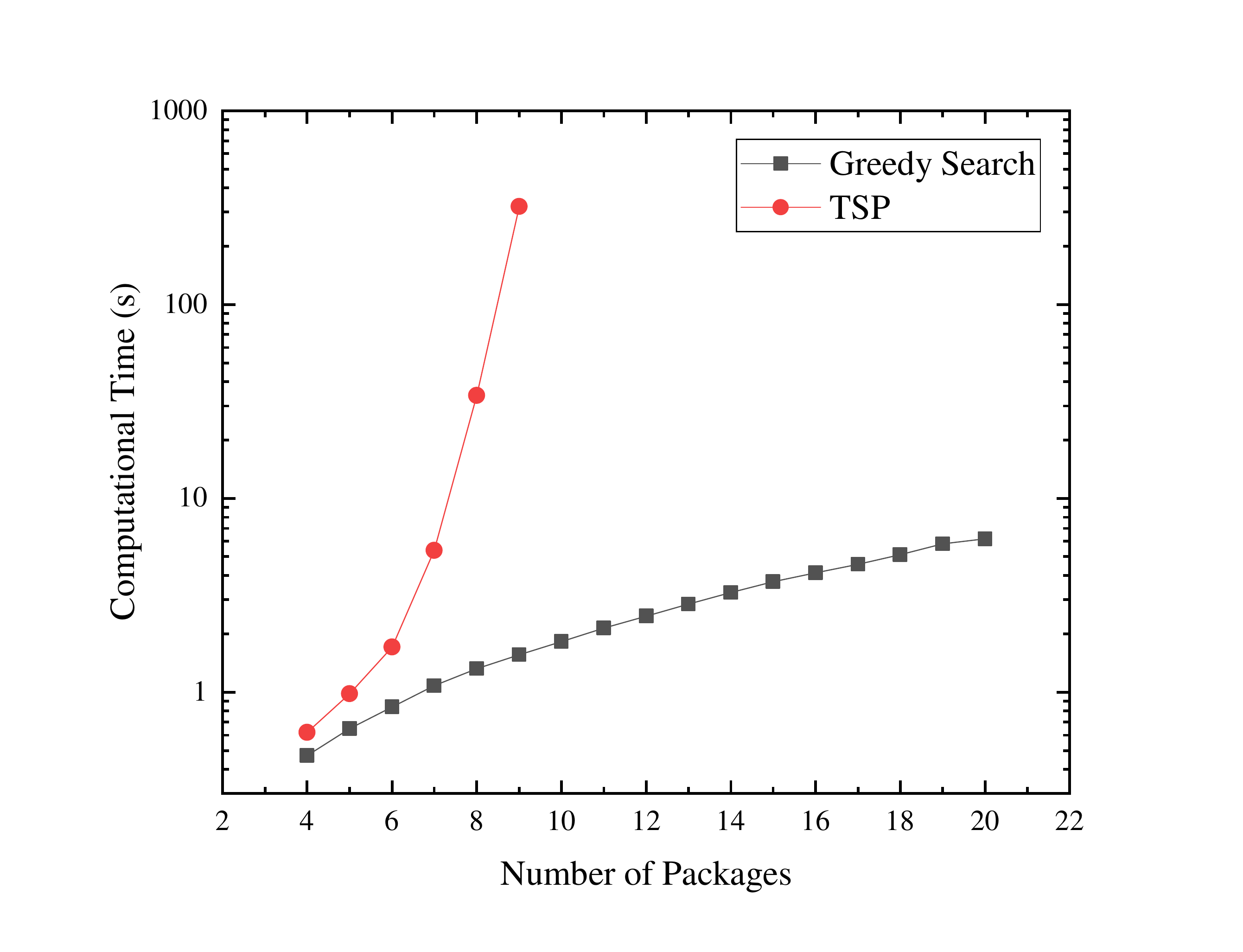}
    \caption{Computational cost between the TSP and greedy search approaches. The computational time is plotted in a log scale due to computational complexity of TSP.}
    \label{fig:compute-time}
\end{figure}

To demonstrate the scalability of our approach, we compared the computational time required to solve TSP and our greedy algorithm.
Specifically, we randomly generated $50$ scenarios for different numbers of packages, from $ M = 4, 5, \dots, 20$.
Due to the computational cost of TSP, we only computed its solution for up to $9$ packages, and the performance of both approaches is demonstrated in Fig. \ref{fig:compute-time}.
Note that the poor performance of the TSP necessitates a log scale on the computational time axis, and this shows the significant computational benefit of using the greedy search algorithm as the number of packages increase.
Even for a small number of packages the TSP solution quickly becomes intractable.
This necessitates sacrificing some optimality to guarantee the scalability of our solution.

\begin{figure}[ht]
    \centering
    \includegraphics[width=\linewidth]{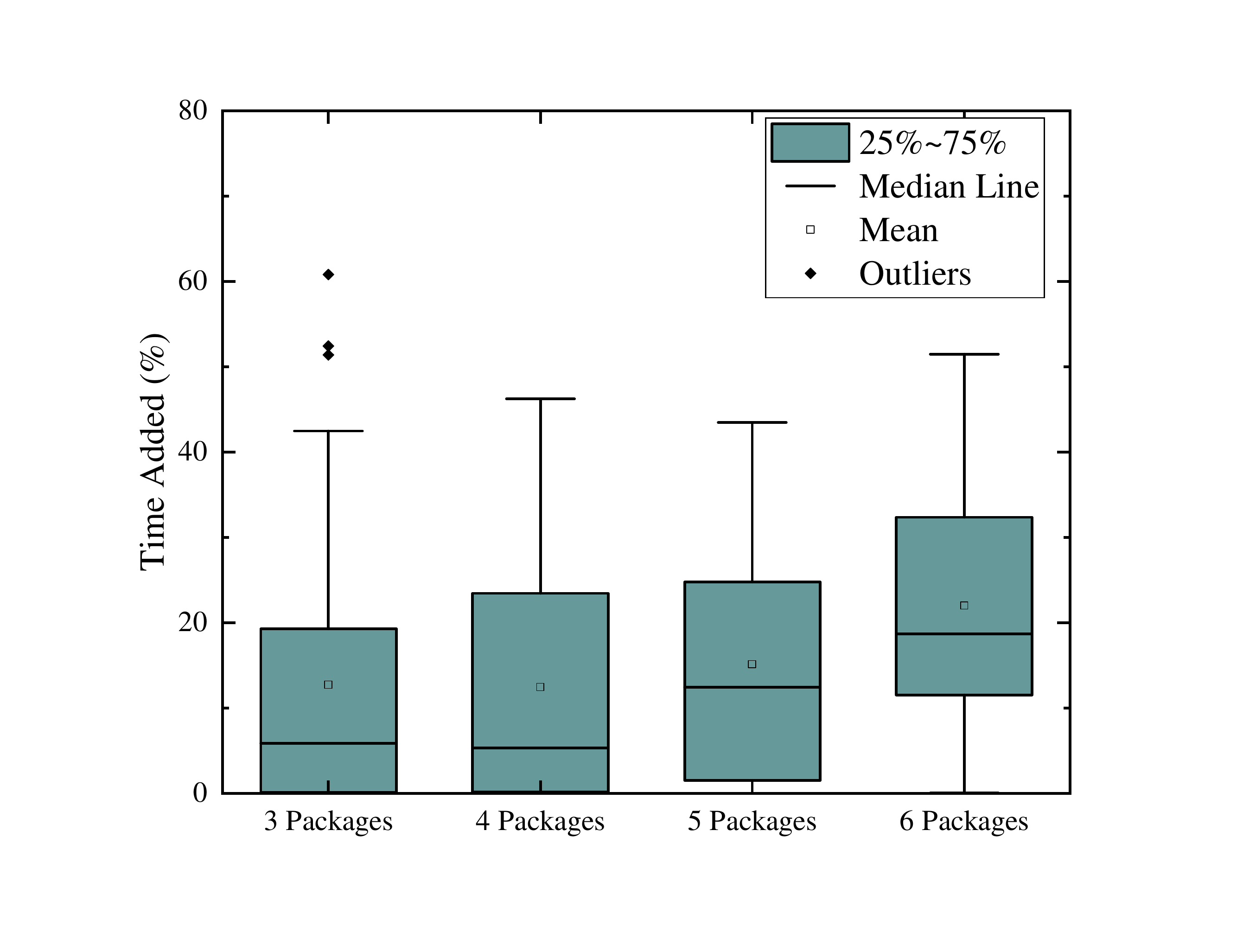}
    \caption{The extra time spent by vehicles to complete their route using the greedy search instead of the TSP solution.}
    \label{fig:time-lost}
\end{figure}

We also conducted $50$ simulations where the package locations were determined randomly for $M=3, 4, 5, 6$ packages.
Figure \ref{fig:time-lost} quantifies the optimality gap of our greedy approach compared to TSP, and shows the percent increase in total travel time compared to the TSP solution.
For low numbers of vehicles, the median time lost is less than 10\%, however, Fig. \ref{fig:time-lost} does demonstrate a consistent growth in the optimality gap.
This gap is less relevant at the number of packages increases, as TSP quickly becomes intractable. 

\section{Simulation and Experiment}\label{sec:simulation}

\begin{figure}[ht]
    \centering
    \includegraphics[width=0.9\linewidth]{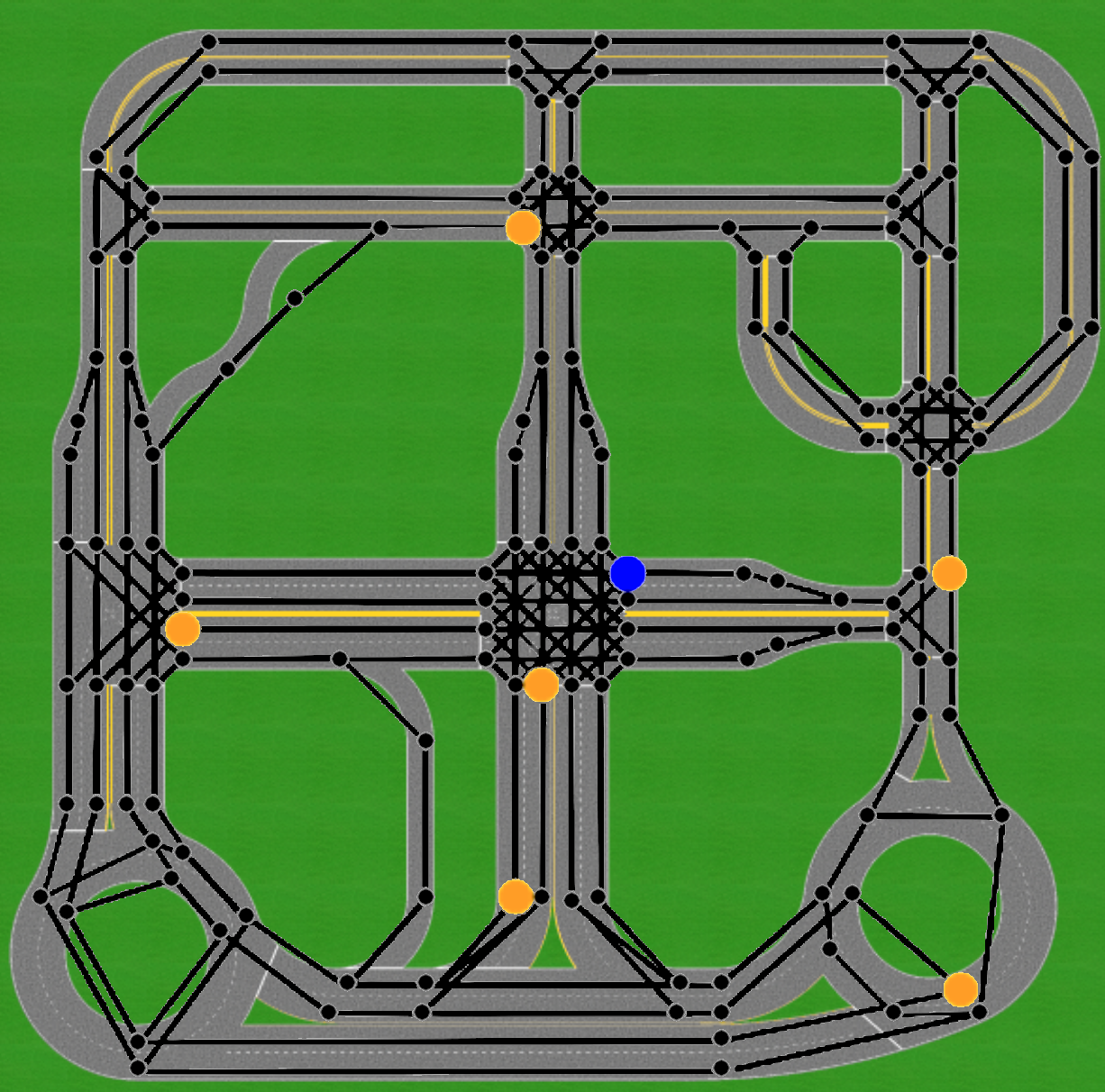}
    \caption{The depot (blue) and package locations (orange) for the 6 package simulation.}
    \label{fig:packages}
\end{figure}

To demonstrate the effectiveness of our approach, we performed a series of simulations based on the urban network present in the Information and Decision Science Lab's Scaled Smart City (IDS$^3$C)\footnote{\url{https://www.youtube.com/watch?v=4x1i3oODn7Q}}. The IDS$^3$C is a $1$:$25$ scaled testbed spanning over $400$ square feet, and it is capable of replicating real-world traffic scenarios using up to $50$ ground and $10$ aerial vehicles; for an overview of the IDS$^3$C and its capabilities see \cite{Beaver2020DemonstrationCity}. 
IDS$^3$C's road network consists of straight lines and arc segments.
To implement our algorithm, we first constructed a graph to represent the IDS$^3$C's road network using network analysis library NetworkX\footnote{\protect\label{note1}NetworkX website: \url{https://networkx.org}} in Python $3.7$.
We constructed the graph's nodes by computing the start and end points of each road segment. 
To eliminate redundant nodes, which were generated due to rounding errors, we combined any two points that were separated by one lane width or less.
The resulting graph of IDS$^3$C consists of $149$ nodes and $220$ edges, and is displayed over a diagram of the IDS$^3$C in Fig. \ref{fig:packages}. 
For each edge in the graph, we defined the cost as the length of the road segment divided by its speed limit.
We considered a speed limit of $50$ km/h for straight roads and $25$ km/h for arc segments.
Figure \ref{fig:packages} also shows the node corresponding to the depot (blue) and six package locations (orange) that we used in the physical experiment, which we discuss in the following section.

\subsection{Simulation Results}

To simulate our last-mile delivery service, a user inputs the number of packages, $M$, and their corresponding delivery locations.
We allow any node except the depot to be a delivery location, as the depot is a fixed node that corresponds to the initial and final destination of all vehicles.
Next, we apply complete-linkage clustering based on the position of each delivery location in $\mathbb{R}^2$ using Scikit-Learn's agglomerative hierarchical clustering package\footnote{Scikit-Learn website: \url{https://scikit-learn.org/stable/}}.
The clustering yields the assignment matrix for every possible number of vehicles.
Then, we iterate through the hierarchy of $1, 2, \dots, N$ clusters and determine the greedy path taken by each vehicle using NetworkX's A$^*$ path finding package\textsuperscript{\normalfont\ref{note1}}.
We used the default setting of no heuristic, thus our solution is equivalent to Dijkstra's algorithm.
Finally, we evaluate the $N$ costs, corresponding to the $1, 2, \dots, N$ vehicles, and select the optimal number of vehicles such that the total cost is minimized for a given value of $\alpha$.
We compute the cost from the total travel time and average delivery time, which correspond to the sum of all edge weights along a vehicle's path and the sum of all edge weights from the origin to the first instant that a vehicle reaches one of its assigned delivery nodes, respectively.
Our pseudocode is detailed in Algorithm \ref{alg:main}.

\begin{algorithm}
\caption{Generate delivery routes for all possible number of vehicles.}\label{alg:main}
\KwIn{Delivery locations of $M$ Packages, Number of available vehicles $N$}
%Randomly select $N$ unique nodes for deliveries\;
$\text{\textit{numVehicles}} \gets 0$ \;
\While{\textit{numVehicles} $\leq$  $N$ }{
 	Group packages into {\it numVehicles} clusters\;
    Assign all packages in each cluster to a vehicle\;
 	\For{each vehicle}{
    	\While{unvisited assigned packages  $>  0$ }{
            Calculate cost of paths to each assigned package delivery point\;
        	Add lowest cost path to the vehicle's path\;
        }
   	}
    $\text{\textit{numVehicles}} \gets \text{\textit{numVehicles}} + 1$ \;
}

\end{algorithm}

%We use total travel time and average delivery time of the delivery routes to compute the cost of using any number of vehicles. Namely, the total travel time for each vehicle is computed by summing all edges' weights in the vehicle's path, while the delivery time for a package is the summation of all edges' weights until the first instance of visiting the delivery location.  

\begin{figure}[ht]
    \centering
    \includegraphics[width=\linewidth]{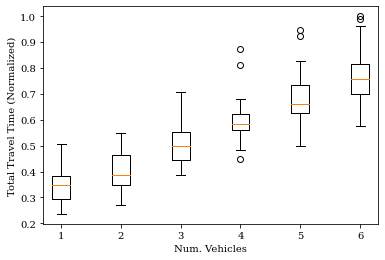}
    \caption{The total time spent by vehicles to complete their route, from 30 tests with 6 packages.}
    \label{fig:whisker-total-time}
\end{figure}

\begin{figure}[ht]
    \centering
    \includegraphics[width=\linewidth]{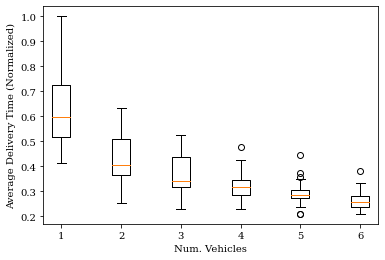}
    \caption{The average delivery time of each package, from 30 tests with 6 packages.}
    \label{fig:whisker-average-time}
\end{figure}

\begin{figure}[ht]
    \centering
    \includegraphics[width=\linewidth]{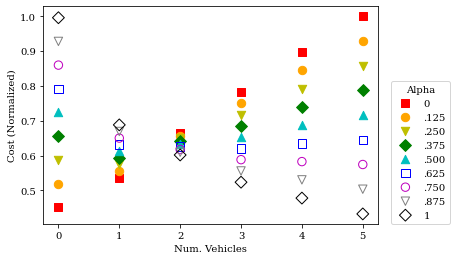}
    \caption{The total cost of the system for different values of $\alpha$, from 30 tests with 6 packages.}
    \label{fig:total-cost}
\end{figure}

To analyze the performance of our last-mile delivery service, we performed $30$ simulations by randomly initializing $M=6$ delivery locations on the network.
The cost components, i.e., total travel time and average delivery time, are presented in Figs. \ref{fig:whisker-total-time} and \ref{fig:whisker-average-time}, respectively.
As expected, Fig. \ref{fig:whisker-total-time} shows that total travel time increases with the number of vehicles, while Fig. \ref{fig:whisker-average-time} demonstrates that the average delivery time decreases as the number of vehicles increases.
These competing objectives present a trade-off between the cost of delivering packages and customer satisfaction.
The total cost, averaged over $30$ random trials, % and using the normalized total and average time cost components, 
is presented in Fig. \ref{fig:total-cost}.
The components of the cost are normalized by dividing them by the maximum cost incurred for each component over all trials and all numbers of vehicles.
% for the $6$ package scenario depicted in Fig. \ref{fig:packages}. 

%We calculated costs of the delivery routes using different values of $\alpha$ and present the averaged results in Fig. \ref{fig:total-cost}. For instance, with $\alpha = 0.5$, the optimal number vehicles is $2$.


\begin{figure}[ht]
    \centering
    \includegraphics[width=\linewidth]{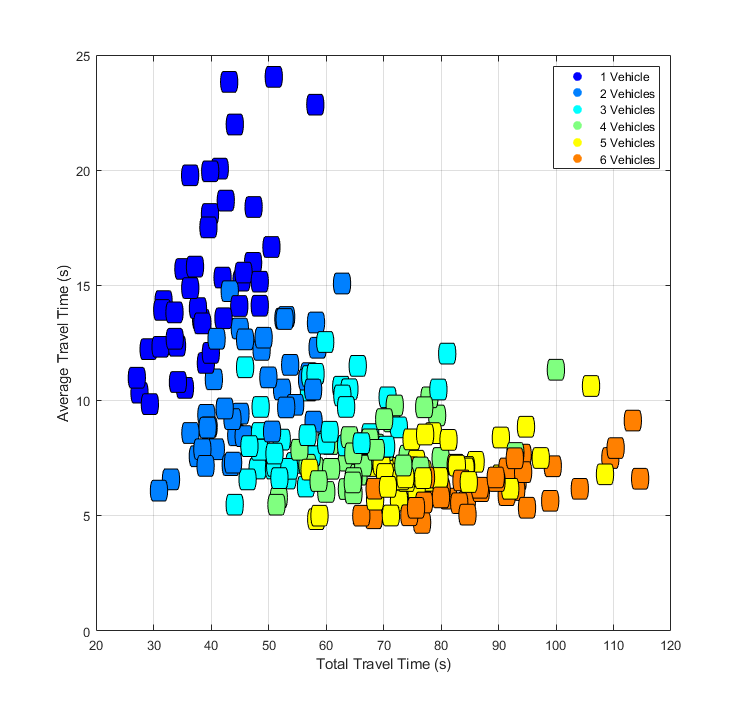}
    \caption{Pareto frontier for $40$ randomized trials with $M=6$ packages and varying the number of vehicles from $1$ to $6$ for each trial.}
    \label{fig:pareto}
\end{figure}

Finally, we performed $40$ randomized simulations with $M=6$ packages to generate a Pareto frontier (Fig. \ref{fig:pareto}).
We note that the cases $N=1$ and $N=6$ are only Pareto-efficient when the value of $\alpha$ is very close to $0$ or $1$, respectively.
The Pareto Frontier shows that the vehicle solutions corresponding to $N = 2, 3, 4$ are the most efficient, with the best Pareto-optimal outcome occurring at approximately $(32, 6)$ for $N=2$ vehicles.

\subsection{Experimental Results} \label{sec:experiment}

%To validate our proposed delivery service we conducted a series of experiments in the IDS$^3$C, a $1$:$25$ scale testbed that uses robotic cars to simulate different traffic scenarios without the safety risks and significant costs of full-scale testbeds.
%The route of each vehicle is fed into a central mainframe computer, which broadcasts real-time trajectory data over wifi to a fleet of autonomous vehicles within the city.
%Each vehicle is equipped with a Raspberry Pi 3B and Pololu Zumo for lane tracking and low-level motor control, respectively.
%For an overview on the IDS$^3$C's hardware and capabilities see \cite{Beaver2019AGeneration,chalaki2020experimental}.

To validate our simulation-based approach to selecting the number of vehicles, we performed an experiment in the IDS$^3$C using the $M=6$ package locations presented in Fig. \ref{fig:packages}.
We applied Algorithm \ref{alg:main} to solve Problems \ref{prb:assignment} and \ref{prb:path}, and used the resulting paths as input for the vehicles in the IDS$^3$C.
Finally, relaxing Assumption \ref{smp:automation}, we determined the control input of each vehicle $i\in\mathcal{N}$ using the Intelligent Driver Model \cite{IDM},
\begin{align} \label{eq:idm} \notag
    a_i =~& a_{\max}\Bigg( 1 - \Bigg(\frac{v_{i}}{v_0}\Bigg)^{\delta} - \Bigg( \frac{s^*}{s_i} \Bigg)^2 \Bigg), \\
    s^* =~& s_0 + v_{i} \theta + \frac{v_{i} \delta \dot{s}_i}{2 \sqrt{(a_{\max}\,a_{\min})}},
\end{align}
where $a_i$, $v_i$, and $s_i$ are the the acceleration, speed, and headway distance of vehicle $i$, respectively, $v_0$ is the speed limit of the road, $s_0$ is the minimum bumper-to-bumper distance, $\theta$ is the desired time headway between vehicles, $a_{\max}$ is the maximum vehicle acceleration, $a_{\min}$ is the maximum (positive) comfortable braking acceleration, and $\delta$ is the exponential factor.
The speed limit of each road segment is determined by the road geometry, denoted $v_s$ for straight line segments, $v_a$ for arc segments, and satisfying $v_s > v_a$.
In addition, the vehicles each stop for $3$ seconds at each delivery node along their route.
The parameters used in the experiment are presented in Table \ref{tab:parameters}.

\begin{table}[ht]
    \centering
    \caption{Parameters used for the intelligent driver model for the physical experiments.}
    \begin{tabular}{ccccccc}
         $\theta$ (s) &  $\delta$ (m) & $a_{\max}$ (m/s$^2$) & $a_{\min}$ (m/s$^2$) & $v_s$ (m/s) & $v_a$ (m/s) \\
         \midrule
            $1.0$    &   $0.06$  &  $5.0$    &   $25.0$    &    $0.5$ & $0.25$
    \end{tabular}
    \label{tab:parameters}
\end{table}

\begin{figure}[ht]
    \centering
    \includegraphics[width=\linewidth]{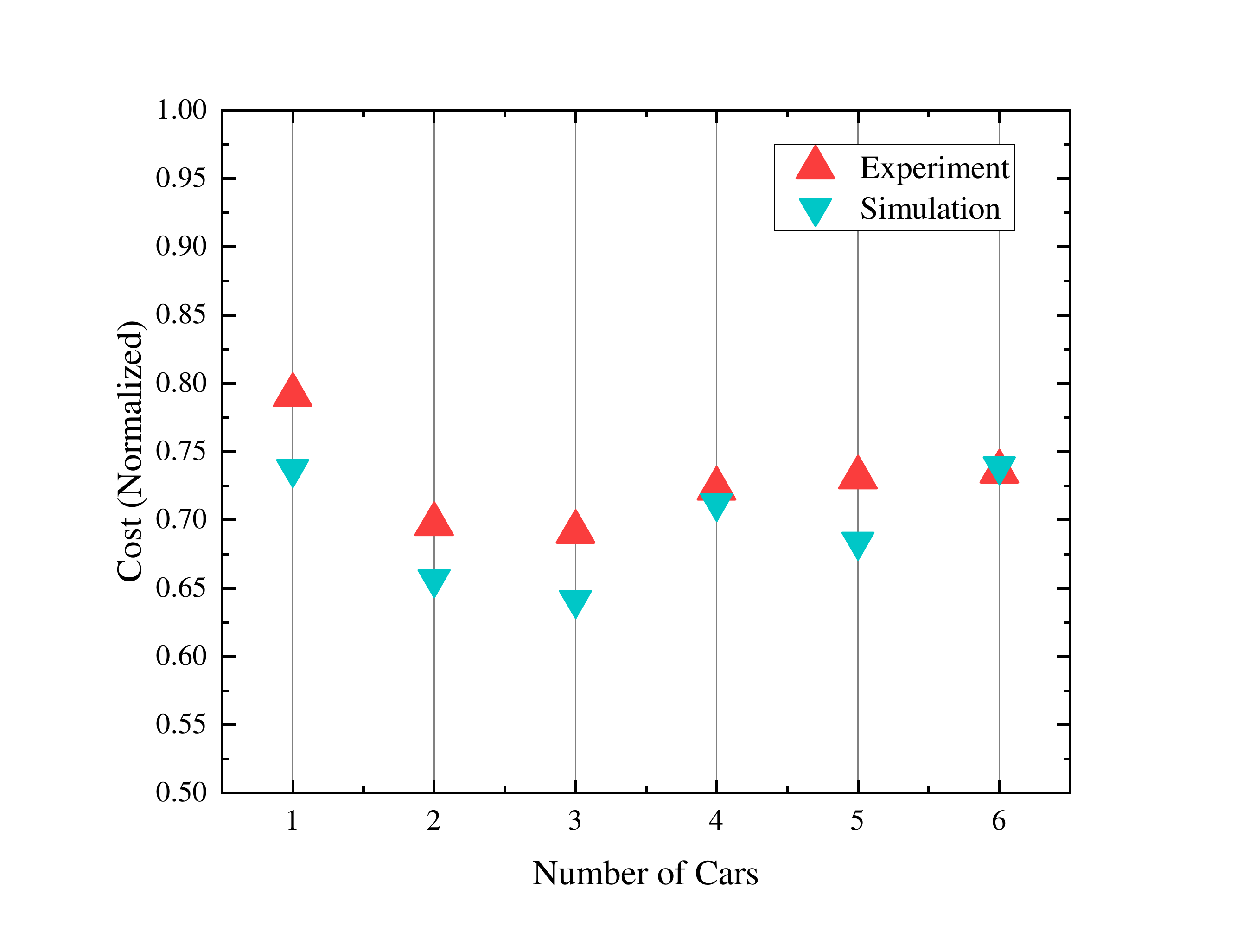}
    \caption{The simulated and actual cost incurred for different numbers of vehicles in the $6$ package delivery scenario with $\alpha = 0 .5$.}
    \label{fig:comparison}
\end{figure}

We performed six experiments, one for each number of vehicles, and computed the customer satisfaction cost \eqref{eq:satisfaction} and total travel time \eqref{eq:final-time} for each case.
In order to compare the simulation and experimental results, we normalized the resulting costs by their maximum value, i.e., we divided each component of the cost by the maximum value that it took in all cases.
The resulting comparison is presented in Fig. \ref{fig:comparison} for $\alpha = 0.5$.
This shows that while the cost of the simulation and experiment differs by at most $10$\%, the simulation correctly predicts $N=3$ as the optimal number of vehicles for this particular scenario.
This is consistent with the Pareto front in Fig. \ref{fig:pareto}.
The simulation consistently under-predicts the total cost of the experiment for $N < 6$, and this is due to the fact that each vehicle stops for $3$ seconds to make a delivery, which increases both the total travel time and subsequent delivery times.
In cases where $N=1$, $N=2,$ and $N=3$, a single vehicle delivers at least $3$ packages, and this results in an average increase of $3$ seconds ($10$\%) per package for that vehicle.
Similarly, when $N=5$ the vehicle that delivers $2$ packages is also assigned to the longest path, which further increases the delivery time.
Finally, for $N=4$, the packages are distributed more evenly between all available vehicles, and in the $6$ vehicle case the delay caused by delivering packages does not affect customer satisfaction.
Experimental results and other supplementary material are available on the paper's website: \url{https://sites.google.com/view/ud-ids-lab/lmds}.

\section{Conclusion}\label{sec:conclusion}

In this paper, we proposed and experimentally validated a heuristic clustering and greedy search algorithm to solve the last-mile delivery problem.
Our approach trades optimality to escape the computational intractability of the TSP coupled to the multi-assignment problem.
We quantified the optimality gap in our approach through random sampling, and developed a fast simulation environment to determine the optimal number of vehicles for our delivery service to use.
Finally, we demonstrated the performance of our delivery service in a scaled testbed using $6$ different experiments.
Future work includes capturing the effect of delays and disturbances in the testbed and simulation, embedding total energy and delivery time constraints in the path generation problem, and enhancing our clustering approach to account for the underlying directed graph that represents the transportation network.

\bibliographystyle{IEEEtran}
\bibliography{IDS_Publications_09022021,References,mendeley}

\begin{thebibliography}{10}
\providecommand{\url}[1]{#1}
\csname url@rmstyle\endcsname
\providecommand{\newblock}{\relax}
\providecommand{\bibinfo}[2]{#2}
\providecommand\BIBentrySTDinterwordspacing{\spaceskip=0pt\relax}
\providecommand\BIBentryALTinterwordstretchfactor{4}
\providecommand\BIBentryALTinterwordspacing{\spaceskip=\fontdimen2\font plus
\BIBentryALTinterwordstretchfactor\fontdimen3\font minus
  \fontdimen4\font\relax}
\providecommand\BIBforeignlanguage[2]{{%
\expandafter\ifx\csname l@#1\endcsname\relax
\typeout{** WARNING: IEEEtran.bst: No hyphenation pattern has been}%
\typeout{** loaded for the language `#1'. Using the pattern for}%
\typeout{** the default language instead.}%
\else
\language=\csname l@#1\endcsname
\fi
#2}}

\bibitem{zhao2019enhanced}
L.~Zhao and A.~A. Malikopoulos, ``Enhanced mobility with connectivity and
  automation: A review of shared autonomous vehicle systems,'' \emph{IEEE
  Intelligent Transportation Systems Magazine}, 2021 (in press).

\bibitem{Hu2016}
M.~Hu and S.~Monhan, ``{US e-Commerce Trends and the Impact on Logistics},''
  \emph{AT Kearney}, 2016.

\bibitem{lin2014survey}
C.~Lin, K.~L. Choy, G.~T. Ho, S.~H. Chung, and H.~Lam, ``Survey of green
  vehicle routing problem: past and future trends,'' \emph{Expert systems with
  applications}, vol.~41, no.~4, pp. 1118--1138, 2014.

\bibitem{Huizing2015}
D.~Huizing, ``Solving the mtsp for fresh food delivery,'' Ph.D. dissertation,
  TU Delft, 2015.

\bibitem{Wang2016}
Y.~Wang, D.~Zhang, Q.~Liu, F.~Shen, and L.~Hay, ``{Towards enhancing the
  last-mile delivery : An effective crowd-tasking model with scalable
  solutions},'' \emph{Transportation Research Part E}, vol.~93, pp. 279--293,
  2016.

\bibitem{yao2015vehicle}
E.~Yao, Z.~Lang, Y.~Yang, and Y.~Zhang, ``Vehicle routing problem solution
  considering minimising fuel consumption,'' \emph{IET Intelligent Transport
  Systems}, vol.~9, no.~5, pp. 523--529, 2015.

\bibitem{jiang2019travelling}
L.~Jiang, H.~Chang, S.~Zhao, J.~Dong, and W.~Lu, ``A travelling salesman
  problem with carbon emission reduction in the last mile delivery,''
  \emph{IEEE Access}, vol.~7, pp. 61\,620--61\,627, 2019.

\bibitem{wang2019car4pac}
F.~Wang, Y.~Zhu, F.~Wang, J.~Liu, X.~Ma, and X.~Fan, ``Car4pac: Last mile
  parcel delivery through intelligent car trip sharing,'' \emph{IEEE
  Transactions on Intelligent Transportation Systems}, vol.~21, no.~10, pp.
  4410--4424, 2019.

\bibitem{osaba2018multi}
E.~Osaba, J.~Del~Ser, A.~J. Nebro, I.~La{\~n}a, M.~N. Bilbao, and J.~J.
  Sanchez-Medina, ``Multi-objective optimization of bike routes for last-mile
  package delivery with drop-offs,'' in \emph{2018 21st International
  Conference on Intelligent Transportation Systems (ITSC)}.\hskip 1em plus
  0.5em minus 0.4em\relax IEEE, 2018, pp. 865--870.

\bibitem{Malikopoulos2016c}
A.~A. Malikopoulos, ``A duality framework for stochastic optimal control of
  complex systems,'' \emph{IEEE Transactions on Automatic Control}, vol.~61,
  no.~10, pp. 2756--2765, 2016.

\bibitem{Baranwal2017}
M.~Baranwal, B.~Roehl, and S.~M. Salapaka, ``Multiple traveling salesmen and
  related problems: A maximum-entropy principle based approach,'' in \emph{2017
  American Control Conference (ACC)}, 2017, pp. 3944--3949.

\bibitem{Bektas2006}
T.~Bektas, ``The multiple traveling salesman problem: an overview of
  formulations and solution procedures,'' \emph{Omega}, vol.~34, no.~3, pp.
  209--219, 2006.

\bibitem{Wang2020}
M.~Wang, T.~Ma, G.~Li, X.~Zhai, and S.~Qiao, ``Ant colony optimization with an
  improved pheromone model for solving mtsp with capacity and time window
  constraint,'' \emph{IEEE Access}, vol.~8, pp. 106\,872--106\,879, 2020.

\bibitem{al2019comparative}
M.~A. Al-Omeer and Z.~H. Ahmed, ``Comparative study of crossover operators for
  the mtsp,'' in \emph{2019 International Conference on Computer and
  Information Sciences (ICCIS)}.\hskip 1em plus 0.5em minus 0.4em\relax IEEE,
  2019, pp. 1--6.

\bibitem{lu2016applying}
Z.~Lu, K.~Zhang, J.~He, and Y.~Niu, ``Applying k-means clustering and genetic
  algorithm for solving mtsp,'' in \emph{International Conference on
  Bio-Inspired Computing: Theories and Applications}.\hskip 1em plus 0.5em
  minus 0.4em\relax Springer, 2016, pp. 278--284.

\bibitem{liu2018research}
C.~Liu and Y.~Zhang, ``Research on mtsp problem based on simulated annealing,''
  in \emph{Proceedings of the 2018 International Conference on Information
  Science and System}, 2018, pp. 283--285.

\bibitem{bramel1995location}
J.~Bramel and D.~Simchi-Levi, ``A location based heuristic for general routing
  problems,'' \emph{Operations research}, vol.~43, no.~4, pp. 649--660, 1995.

\bibitem{trigui2017fl}
S.~Trigui, O.~Cheikhrouhou, A.~Koubaa, U.~Baroudi, and H.~Youssef, ``Fl-mtsp: a
  fuzzy logic approach to solve the multi-objective multiple traveling salesman
  problem for multi-robot systems,'' \emph{Soft Computing}, vol.~21, no.~24,
  pp. 7351--7362, 2017.

\bibitem{Tong2019}
P.~Tong, W.~Du, M.~Li, J.~Huang, W.~Wang, and Z.~Qin, ``Last-mile school
  shuttle planning with crowdsensed student trajectories,'' \emph{IEEE
  Transactions on Intelligent Transportation Systems}, vol.~22, no.~1, pp.
  293--306, 2021.

\bibitem{naumov2021identifying}
V.~Naumov and M.~Pawlu{\'s}, ``Identifying the optimal packing and routing to
  improve last-mile delivery using cargo bicycles,'' \emph{Energies}, vol.~14,
  no.~14, p. 4132, 2021.

\bibitem{remer2019multi}
B.~Remer and A.~A. Malikopoulos, ``The multi-objective dynamic traveling
  salesman problem: Last mile delivery with unmanned aerial vehicles
  assistance,'' in \emph{2019 American Control Conference (ACC)}.\hskip 1em
  plus 0.5em minus 0.4em\relax IEEE, 2019, pp. 5304--5309.

\bibitem{di2020trucks}
L.~Di~Puglia~Pugliese, G.~Macrina, and F.~Guerriero, ``Trucks and drones
  cooperation in the last-mile delivery process,'' \emph{Networks}, 2020.

\bibitem{Beaver2019AGeneration}
L.~E. Beaver and A.~A. Malikopoulos, ``{A Decentralized Control Framework for
  Energy-Optimal Goal Assignment and Trajectory Generation},'' in \emph{IEEE
  58th Conference on Decision and Control}, 2019, pp. 879--884.

\bibitem{bang2021energy}
H.~Bang, L.~E. Beaver, and A.~A. Malikopoulos, ``Energy-optimal goal assignment
  of multi-agent system with goal trajectories in polynomials,'' in \emph{2021
  29th Mediterranean Conference on Control and Automation (MED)}.\hskip 1em
  plus 0.5em minus 0.4em\relax IEEE, 2021, pp. 1228--1233.

\bibitem{Mahbub2019ACC}
A.~M.~I. Mahbub, L.~Zhao, D.~Assanis, and A.~A. Malikopoulos, ``{Energy-Optimal
  Coordination of Connected and Automated Vehicles at Multiple
  Intersections},'' in \emph{Proceedings of 2019 American Control Conference},
  2019, pp. 2664--2669.

\bibitem{chalaki2020TCST}
B.~Chalaki and A.~A. Malikopoulos, ``Optimal control of connected and automated
  vehicles at multiple adjacent intersections,'' \emph{IEEE Transactions on
  Control Systems Technology (in press)}, 2021.

\bibitem{dawyndt2005complete}
P.~Dawyndt, H.~De~Meyer, and B.~De~Baets, ``The complete linkage clustering
  algorithm revisited,'' \emph{Soft Computing}, vol.~9, no.~5, pp. 385--392,
  2005.

\bibitem{Beaver2020DemonstrationCity}
L.~E. Beaver, B.~Chalaki, A.~M. Mahbub, L.~Zhao, R.~Zayas, and A.~A.
  Malikopoulos, ``{Demonstration of a Time-Efficient Mobility System Using a
  Scaled Smart City},'' \emph{Vehicle System Dynamics}, vol.~58, no.~5, pp.
  787--804, 2020.

\bibitem{IDM}
M.~Treiber, A.~Hennecke, and D.~Helbing, ``Congested traffic states in
  empirical observations and microscopic simulations,'' \emph{Phys. Rev. E},
  vol.~62, pp. 1805--1824, Aug 2000.

\end{thebibliography}

\end{document}